\documentclass[%
reprint,
superscriptaddress,
 amsmath,amssymb,
 aps, pra,
]{revtex4-2}
\usepackage{upgreek}
\usepackage{graphicx}
\usepackage{dcolumn}
\usepackage{bm}
\usepackage{hyperref}
\usepackage[mathlines]{lineno}

\begin{document}


\title{Controlling emitter-field coupling in waveguides with a nanomechanical phase shifter}
\author{Celeste Qvotrup}
\thanks{These authors contributed equally.}
\author{Ying Wang $^\dag$}
\thanks{These authors contributed equally.}
\author{Marcus Albrechtsen}
\author{Rodrigo Thomas}
\author{Zhe Liu}
\affiliation{Center for Hybrid Quantum Networks (Hy-Q), Niels Bohr Institute, University of Copenhagen, Blegdamsvej 17, Copenhagen, 2100, Denmark.}
\author{Sven Scholz}
\affiliation{Lehrstuhl f{\"u}r Angewandte Festk{\"o}rperphysik, Ruhr-Universit{\"a}t Bochum, Universit{\"a}tsstrasse 150, D-44780 Bochum, Germany}
\author{Arne Ludwig}
\affiliation{Lehrstuhl f{\"u}r Angewandte Festk{\"o}rperphysik, Ruhr-Universit{\"a}t Bochum, Universit{\"a}tsstrasse 150, D-44780 Bochum, Germany}

\author{Leonardo Midolo}

\thanks{Email to: ying.wang@nbi.ku.dk; midolo@nbi.ku.dk}
\affiliation{Center for Hybrid Quantum Networks (Hy-Q), Niels Bohr Institute, University of Copenhagen, Blegdamsvej 17, Copenhagen, 2100, Denmark.}

\date{\today}

\begin{abstract}
The ability to control light-matter interfaces with solid-state photon emitters is a major requirement for the development of quantum photonic integrated circuits. We demonstrate controllable coupling between a quantum dot and an optical mode in a dielectric waveguide using a nano-opto-electro-mechanical phase shifter and a photonic crystal mirror. By controlling the phase, we induce a virtual displacement of the mirror that modifies the local density of states at the position of the emitter, thereby enhancing or suppressing spontaneous emission without an optical cavity. We observe a broadband tuning of the spontaneous emission rate and a modulation of the intensity emitted by the quantum dot in the waveguide. The method reported here could be employed to optimize the emitter-field interaction between quantum dots in-operando, by maximizing a single-photon source generation rate or adjusting its lifetime, as well as a characterization tool for the direct measurement of emitter-photon cooperativity. 
\end{abstract}

\maketitle


\section{\label{sec:level1}Introduction}
Solid-state quantum emitters are essential resources for the realization of quantum photonic integrated circuits with near-deterministic single-photon sources and spin-photon interfaces \cite{wang2020integrated,uppu2021quantum,simmons2024scalable}. A known issue towards scaling up the number of emitters that can be simultaneously controlled and addressed in a chip, is to reliably achieve optimal spatial alignment between optical modes and transition dipoles to maximize the local density of optical states (LDOS). 
A widely-used approach to circumvent spectral and spatial randomness, is to pre-characterize  the emitters prior to device fabrication, to identify suitable candidates that can be subsequently aligned to nanostructures \cite{dousse2008controlled,gschrey2015highly,pregnolato2020deterministic}. These methods rely on ultra-precise fabrication and imaging methods to guarantee precise spatial alignment and poses constraints on the design of photonic integrated circuits. An alternative approach is to implement reconfigurable nanophotonic devices to adjust the emitter-field coupling via an external control. The majority of work has focused on methods to adjust the spectral alignment between nanocavities, such as micropillars and photonic crystal cavities, and emitters. A wide range of tuning mechanisms including strain tuning \cite{moczala2019strain}, Stark effect \cite{faraon2010fast}, mechanical systems \cite{midolo2012spontaneous}, and thermo-optic tuning \cite{elshaari2017chip} have been demonstrated. However, the in-operando spatial alignment, (i.e., adjusting the relative position between and emitter and an optical mode via an external control) is significantly less explored.

\begin{figure}[ht]
\includegraphics{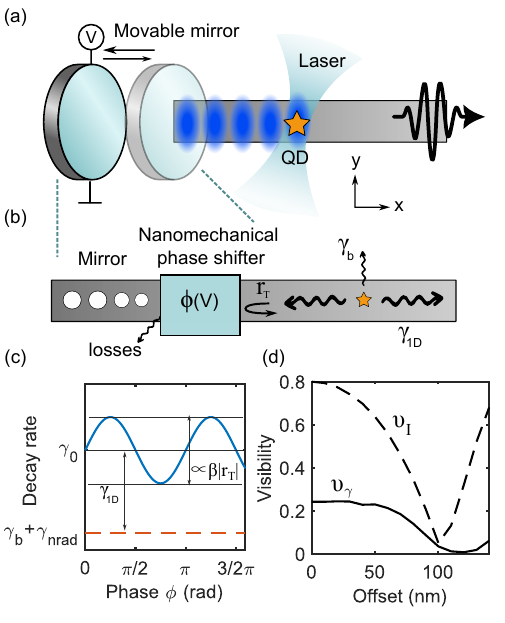}
\caption{\label{fig:figure_1} One-dimensional implementation of emitter-field coupling control. (a) Schematic of a waveguide with a movable mirror on one end, enabling the control of the local density of optical states at the location of a distant quantum dot (QD). (b) Experimental realization with a fixed photonic crystal mirror and a tunable voltage-controlled nanomechanical phase shifter. (c) Expected modulation of decay rate as a function of the phase in the presence of finite mirror reflectivity $r_\text{T}$, leak rate outside the waveguide  $\gamma_\text{b}$ and non-radiative decay rate $\gamma_\text{nrad}$. (d) Calculated visibilities for collected intensity ($\nu_\text{I}$) and decay rate ($\nu_\gamma$) as a function of the offset from the waveguide center assuming $r_\text{T}=0.5$ and above-band excitation. }
\end{figure}

In this work, we investigate quantum emitters in one-sided one-dimensional (1D) waveguides and demonstrate a very broadband coupling and a controllable LDOS by adding a tunable phase shifter in the waveguide. The phase shifter controls the interference between the emitted light and its reflection off a distant mirror, thereby allowing a remote control of spontaneous emission even in the absence of an optical cavity.
The experiment described above, is schematically shown in Figure~\ref{fig:figure_1}(a), which is essentially the 1D version of Drexhage's experiment originally performed with fluorescent molecules \cite{drexhage1970influence,frimmer2013spontaneous}. Here, the 1D configuration allows the mirror to be placed at considerably longer distances (in this experiment $L \sim 70 \lambda$ or $\sim30$ $\upmu \text{m}$) than the equivalent experiments in free-space. We employed self-assembled quantum dots (QDs) in GaAs waveguides, and implemented an ultra-compact nano-opto-electro-mechanical system (NOEMS) \cite{midolo2018nano} to tune the phase in slot-mode waveguides with an external applied voltage (Fig.~\ref{fig:figure_1}(b)). We achieve a 50\% lifetime modulation between $(1.00 \pm 0.08)$ ns when enhanced by the mirror and $(1.53 \pm 0.08)$ ns when suppressed, limited by the phase shifter losses and the non-resonant excitation scheme employed in the experiment.  The design, fabrication and characterization of the NOEMS phase shifter is discussed in a companion paper \cite{companion}. 

\section{\label{sec:theory} Vacuum field control}
The virtual motion of the mirror manifests itself in two different ways: 1) an interference effect in the counts at the collection of the grating, and 2) a modulation of the spontaneous emission rate of the emitter. The first effect is due to the standing wave caused by the wave reflected at the mirror, given by
\begin{equation}
\frac{I(\phi)}{I_0} = \frac{1}{2}\left(1 + |r_\text{T}|^2 \pm 2|r_\text{T}|\cos\left(2\phi+\theta\right)\right),
\end{equation}
where $I_0$ is the radiated intensity in the absence of mirrors and $\theta=2kL$ is a fixed phase depending on the optical distance between the emitter and the mirror. $r_\text{T}$ is the lumped reflectivity of the mirror that combines the transmittivity of the NOEMS phase shifter $t_\phi$, the waveguide transmittivity $t_\text{wg}$, and photonic crystal mirror reflectivity $r_\text{M}$, i.e. $r_\text{T} = |t_\phi^2 t_\text{wg}^2 r_\text{M}|\exp(i 2\phi)$.
The positive and negative sign is used for $x$- and $y$-polarized dipoles, respectively.
The total  visibility, defined as $\nu_\text{I} = (I_\text{max}-I_\text{min})/(I_\text{max}+I_\text{min})$, is thus:
\begin{equation}
\nu_\text{I} = \frac{2|r_\text{T}|}{1 + |r_\text{T}|^2}.
\end{equation}

The lifetime is controlled by the rate of spontaneous emission in the waveguide $\gamma_0$ and the combined decay rate due to coupling to leaky modes $\gamma_\text{b}$ and non-radiative decay processes $\gamma_\text{nrad}$. Combined, they define a $\beta$-factor, i.e. $\beta _0= \gamma_0/(\gamma_0+\gamma_\text{b})=\gamma_0/\Gamma_0$ where $\Gamma_0 \equiv \gamma_0+\gamma_\text{b}$. We omit the non-radiative contribution in the following, as typical values of $\gamma_\text{nrad}\simeq 0.1$ ns$^{-1} \ll\gamma_\text{rad}$ \cite{johansen2010probing}. The total decay rate $\Gamma_\phi(\phi) $ in the presence of a virtual mirror displacement can be calculated from Green's function using a mirror dipole (see Appendix A), resulting in:
\begin{equation}\label{eq:ldoschange}
\Gamma_\phi(\phi) = \gamma_{\phi}(\phi) + \gamma_b = \gamma_{0}\left(1 \pm |r_\text{T}|\cos(2\phi+\theta)\right) + \gamma_b,
\end{equation}
where $\gamma_{0}$ is the (1D) decay rate in the absence of a mirror, with corresponding coupling $\beta_0$. The expected response and relevant parameters are shown in Fig.~\ref{fig:figure_1}(c).
We argue that the background coupling $\gamma_\text{b}$ to leaky radiation modes is independent of the phase $\phi$, allowing us to derive a total spontaneous emission rate without any prior knowledge on $\beta_0$. In fact, this is usually done in most works by numerical simulations or comparison to other emitters lifetime under the assumption that all dots have identical oscillator strength \cite{arcari2014near}. With a variable phase, however, we can compare the visibility of the collected emission signal to the variation in lifetime rate and use it to extract $\beta_0$. The resulting visibility of the total decay rate is
\begin{equation}
\nu_\gamma = \beta_0 |r_\text{T}|.
\end{equation}
Thus, in principle, by measuring both $\nu_I$ and $\nu_\gamma$ for a single transition dipole, it is possible to characterize simultaneously the combined losses of the mirror and phase shifter, as well as the QD $\beta$-factor.
Yet, one significant challenge in this scheme is that the equations above hold strictly for single dipole orientations in single-mode waveguides. In this work, performed without gated QDs in undoped membranes, we have employed non-resonant excitation schemes, which equally populate both dipolar transition, causing an effective decay rate given by the average of the two dipole rates $\Gamma_{tot}(\phi) = \frac{1}{2}(\Gamma_{x\phi}(\phi) + \Gamma_{y\phi}(\phi)$) \cite{stepanov2015quantum}.  
The corresponding visibility is thus reduced to: 
\begin{equation}
\nu_\gamma = \left|\beta_{x0}\frac{\Gamma_{x0}}{\Gamma_{x0}+\Gamma_{y0}}-\beta_{y0} \frac{\Gamma_{y0}}{\Gamma_{x0}+\Gamma_{y0}}\right| |r_\text{T}|.
\end{equation}
The radiative rate for an $X-$ or $Y-$oriented dipole differ only slightly as a function of the lateral offset, i.e. $\Gamma_{x0}\simeq\Gamma_{y0}$, thus for QDs located in the waveguide center, where $\beta_{x0}\simeq0$, a good approximation is $\nu_\gamma \simeq \frac{1}{2}\beta_{y0} |r_\text{T}|$.

The count intensity is also modified due to the combined modulation of the two dipoles (cf. Appendix A), resulting in:
\begin{equation}
\label{eq:vis_counts_full}
    \nu_\text{I} = \frac{2|r_\text{T}|}{1+|r_\text{T}|^2} \left|\frac{\left|e_y(y_0)\right|^2-\left|e_x(y_0)\right|^2}{\left|e_y(y_0)\right|^2+\left|e_x(y_0)\right|^2}\right|
\end{equation}
where $e_x$ and $e_y$ are the normalized electric field eigenmodes of the rectangular waveguide. Both visibilities have been computed numerically using finite element simulation of a rectangular waveguide and plotted in Fig.~\ref{fig:figure_1}(d) for a mirror reflectivity of $|r_\text{T}| = 0.5$, which is the estimated value for the devices used in this work. 

\section{Experimental control of spontaneous emission}
\begin{figure}[ht]
\includegraphics{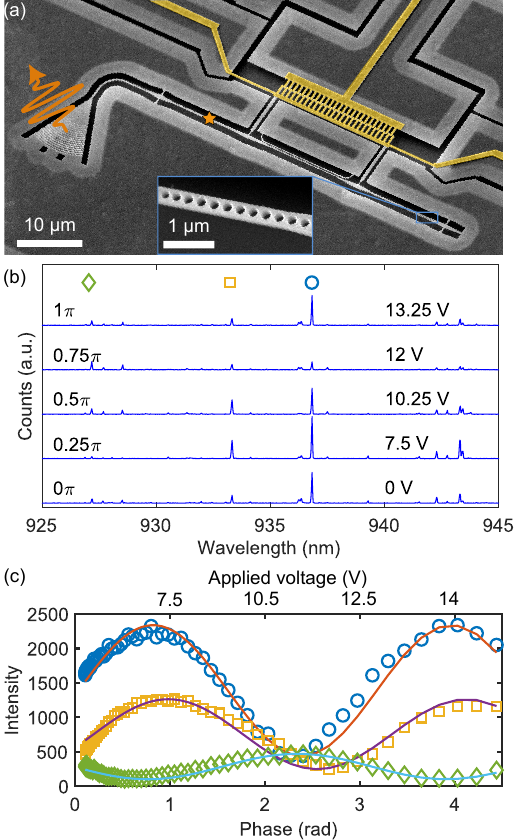}
\caption{Experimental realization of movable mirrors. (a) Scanning electron microscope image of the nanophotonic device used in the experiments. A slot-waveguide phase shifter coupled to an electrostatic actuator (highlighted in yellow) is placed before a 1D photonic crystal mirror (inset). The QDs are excited in the area marked by the star and light is collected via a grating coupler. (b) Spectra acquired under above-band excitation at different voltages and corresponding phase shift, showing several excitonic lines being modulated in intensity. (c) Integrated intensity from three lines in (b) as a function of voltage. Sinusoidal curves (solid lines) are overlaid for comparison.}
\label{fig:figure_2}
\end{figure}

The integrated circuit featuring a photonic crystal mirror, a phase shifter, and an outcoupling grating was fabricated in an undoped GaAs membrane with medium-density ($\sim 10$ $\upmu$m$^{-2}$ self-assembled InAs QDs). A scanning electron microscope image of the device is shown in Figure~\ref{fig:figure_2}(a). The photonic crystal mirror is made by etching an array of 12 consecutive holes, spaced by 265 nm, with a radius of 70 nm. From numerical simulations, the mirror provides $>90$\% reflectivity over a wide band of 900-1000 nm.   The 1D suspended waveguide is 300 nm wide, ensuring single-mode operation for transverse electric modes. From previous works \cite{papon2019nanomechanical}, such suspended waveguide provide $>$ 7.5 dB/mm loss due to sidewall roughness, resulting in a waveguide efficiency (over the QD-mirror distance of $30$ $\upmu $m of $|t_\text{wg}|^2\sim 0.9$. 
Typical insertion loss of the NOEMS phase shifter is approximately $-2.5$ dB ($|t_\phi|^2\simeq0.55$), resulting in a total combined mirror reflectivity of $|r_\text{T}|\simeq0.5$.

The sample is wire-bonded to a printed circuit board and mounted inside a closed-cycle cryostat operating at 4 K. We excite the QDs with an above-band pulsed laser at 800 nm. The emitted light is collected from the shallow-etched grating and analyzed in a spectrometer. The grating is designed to collect only the fundamental transverse electric mode (TE0) while rejecting higher-order odd modes. For lifetime measurements, we included a  tunable filter (bandwidth of 0.5 nm) to select individual excitonic transitions and performed time-resolved spectroscopy with superconducting nanowire single-photon detectors (SNSPD). All measurements are performed while sweeping the voltage that controls the phase shifter.

Figure~\ref{fig:figure_2}(b) shows the spectrum as a function of the applied voltage and the corresponding phase shift.  Several QD emission lines are modulated in intensity without any observable spectral tuning. The location of the excitation spot is indicated approximately in Fig.~\ref{fig:figure_2}(a). A similar response is observed on different excitation spots between the phase shifter and the collection grating. The total intensity of QD emission is plotted in Fig.\ref{fig:figure_2}(c) for three representative QD lines. The phase-voltage mapping used in the plot is reconstructed using the bright emission line at 937 nm (blue circles). All the lines exhibit a clear sinusoidal dependence with a different phase offset, originating from the different position along the waveguide, but very similar visibility $\nu_\text{I}= (0.67 \pm 0.05)$.  From the visibility, we conclude that $|r_\text{T}| > 0.4$ (where the lowest bound is extracted assuming a perfectly-centered QD), compatible with the estimation of device losses.

\begin{figure}[ht]
\includegraphics{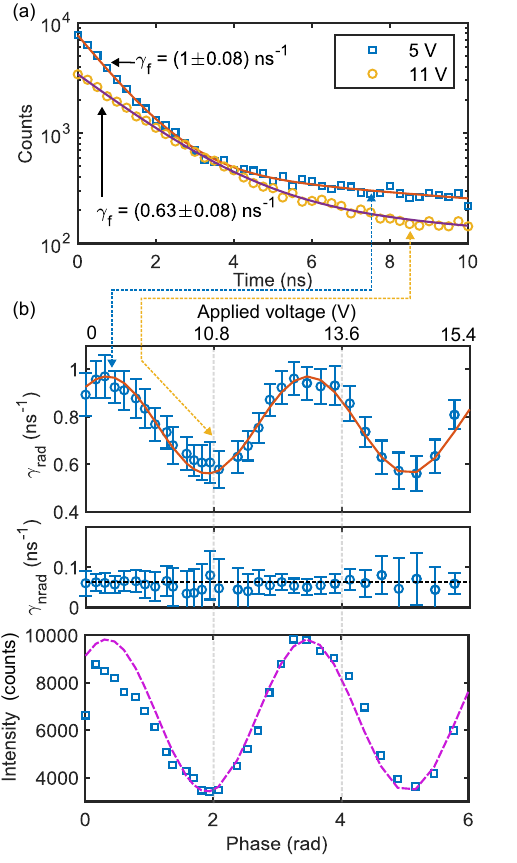}
\caption{Control of spontaneous emission rate. (a) Time-resolved photoluminescence of a QD for two applied voltages, corresponding to the maximum and minimum observed decay rate. The solid lines are bi-exponential fits to the curves. The data is not normalized. (b) From top to bottom, phase-dependent spontaneous decay rates for the radiative ($\gamma_\text{rad}$), non-radiative ($\gamma_\text{nrad}$ contribution, and the total intensity modulation. Solid lines are fits to the data.}
\label{fig:figure_3}
\end{figure}

The time-resolved fluorescence of a single QD line at $\lambda = 923.25$ nm is shown in Figure~\ref{fig:figure_3}(a) for two applied voltages, corresponding to the highest and lowest intensity. Two distinct decay rates are observed, as commonly reported for neutral excitons. We fit the decay rate with two exponential functions as described in ref. \cite{johansen2010probing}.
Here, the fast decay rate $\gamma_f$ is due to the bright transition, while the slow rate $\gamma_s$ probes the non-radiative decay rate from the dark exciton. A slow ($>1$ $\upmu$s, typically) spin-flip rate couples the bright and dark transitions, allowing us to accurately extract the radiative decay rate from $\gamma_\text{rad} = \gamma_\text{f} - \gamma_\text{s}$ and the non-radiative rate from $\gamma_\text{nrad} \simeq \gamma_\text{s}$. The results are plotted in Fig.\ref{fig:figure_3}(b) as a function of voltage and phase, together with the total intensity. It is evident, that both lifetime and intensity follow the same oscillatory response, with a faster (slower) decay rate at the corresponding highest (minimum) emission intensity. The lifetime is modulated between $(1.00 \pm 0.08)$ ns, when enhanced by the mirror and $(1.53 \pm 0.08)$ ns, when suppressed (i.e., over 50\% modulation), with a total visibility of $\nu_\gamma = (0.27 \pm 0.04)$. The visibility of the intensity is $\nu_\text{I} = (0.48\pm0.01)$, suggesting that the QD is likely offset by approximately 60 nm laterally and $|r_\text{T}|\sim0.6$. We report in Table \ref{tab:table_1}, the measurements of five more quantum dot lines, all exhibiting larger count visibility but overall smaller modulation of lifetime. 

\begin{table}
\caption{\label{tab:table_1}Measured lifetime modulation for different QDs in the same device.}
\begin{ruledtabular}
\begin{tabular}{llllll}
QD & $\lambda$ (nm) & $\gamma_\text{max}$ (ns$^{-1}$) & $\gamma_\text{min}$ (ns$^{-1}$) & $\nu_\gamma$ & $\nu_\text{I}$\\
1 & 923.45 &1.00$\pm0.08$ &0.63$\pm0.08$ & 0.27$\pm 0.04$ & 0.42\\
2 & 925.82 &1.14$\pm0.01$ &0.77$\pm0.01$ & 0.15$\pm0.01$ & 0.64\\
3 & 936.82 &0.96$\pm0.01$ & 0.58$\pm0.07$& 0.18$\pm0.03$ & 0.65\\
4 & 940.21 &0.83$\pm0.01$ &0.60$\pm0.05$ & 0.17$\pm0.01$ & 0.70\\
5 & 942.74 &1.29$\pm0.03$ &1.07$\pm0.05$ & 0.08$\pm0.01$ & 0.62\\
6 & 945.98 &1.31$\pm0.01$ & 0.90$\pm0.01$& 0.18$\pm0.01$ & 0.83\\
\end{tabular}
\end{ruledtabular}
\end{table}

\section{Discussion}
In our experiment, carried out with above-band laser excitation, we do not distinguish the two $x-$ and $y-$dipole as the fine structure splitting of $<5$ GHz is too small to be resolved in the spectrometer. Hence, we do not have a direct way to measure the losses in our system and the ideal $\beta-$factor in waveguides. Nevertheless, from the observed visibilities, we can however conclude that the overall reflectivity of the mirror is somewhere between 0.4 and 0.6, and that the QDs are likely distributed with different lateral offsets, potentially all the way to the edge of the waveguide.

Our work shows that individual quantum emitters can be spatially aligned to optical modes using a simple integrated phase shifter and a mirror, enabling the control of the emission intensity and lifetime. 
We note that with the method presented in this work combined with resonant excitation schemes, it would be possible to directly measure $\beta$ for each emitter and potentially deduce the lateral offset. This would not only allow the direct control of spontaneous emission or emitter-photon coupling, but also could serve as a characterization tool to determine the best quantum emitter among randomly-distributed QD ensembles. Controlling light-matter interaction is key in scaling up to large number of emitters and compensate for the intrinsic randomness of self-assembled QDs \cite{elshaari2017chip,dusanowski2023chip,papon2023independent}. Future work will explore the applications of on-chip phase shifters to control the radiative coupling between distant emitters, for example to switch between dispersive and dissipative collective decay dynamics \cite{tiranov2023collective} in a controllable manner and potentially scaling up to large networks of quantum dots. 

\begin{acknowledgments}
The authors acknowledge Anders S. S{\o}rensen and P. Lodahl, for insightful discussions and Andreas Wieck for providing the equipment to grow the nanostructures. 
We acknowledge funding from the European Research Council (ERC) under the European
Union’s Horizon 2020 research and innovation program (No. 949043, NANOMEQ), the Danish National
Research Foundation (Center of Excellence “Hy-Q,” grant number DNRF139),   Styrelsen for Forskning og
Innovation (FI) (5072- 00016B QUANTECH), BMBF QR. N project 16KIS2200, QUANTERA BMBF EQSOTIC project 16KIS2061, as well as DFG excellence cluster ML4Q project EXC 2004/1. 
\end{acknowledgments}

\appendix

\section{Control of spontaneous emission}
To describe the dynamics of the interaction between QDs and the optical modes, we consider the layout shown in figure \ref{fig:figure_1}(b). The waveguide is oriented along the $x$-axis, and the electric dipole source (oriented along $x$ or $y$) is located at the origin $x=0$, $z=0$ (center of the waveguide along its thickness) but potentially offset along $y$ at a position $y_0$. The mirror is located at a distance $|x_m| = L$ and it is characterized by a single reflectivity $r_\text{T} = |r_\text{T}|e^{i2\phi}$ and variable phase $\phi$. 
With this layout, the electric field at the position of the emitter is given by
\begin{equation}
     \mathbf{E}(0,y_0,z_0) =  \mathbf{\bar{E}}(y_0,z_0) +
 |r|\mathbf{\bar{E}}^\ast(y_0,z_0) e^{-i2(kL +\phi)}
\end{equation}
where the first term is the field emitted by the source and the second term is the reflected wave from the mirror. We denote with $\mathbf{\bar{E}(y,z)}$ the transverse eigenfunctions of the waveguide, which can be obtained solving Helmoltz' equation numerically for the specific waveguide shape. In rectangular waveguides the transverse-electric modes at $z=0$, i.e. where the QDs are located, a good approximation for the transverse eigenfunction is $\mathbf{\bar{E}} = \left(i e_x(y), e_y (y),0\right)$, with $e_x$ and $e_y$ purely real functions.  
The light intensity radiated by the an $x-$dipole at the origin is thus:
\begin{equation}
    I_x = \frac{cn\varepsilon_0}{2}|e_x(y_0)|^2\left[1 + |r_\text{T}|^2 + 2|r_\text{T}|\cos(2\phi+2kL)\right],
\end{equation}
whereas for a $y-$dipole:
\begin{equation}
    I_y = \frac{cn\varepsilon_0}{2}|e_y(y_0)|^2\left[1 + |r_\text{T}|^2 - 2|r_\text{T}|\cos(2\phi+2kL)\right].
\end{equation}
Thus, as expected, the radiated power oscillates sinusoidally with the relative distance from the mirror $L$ and the variable phase $\phi$. The visibility of the intensity collected at the grating is independent of the lateral offset $y_0$ for a given dipole orientation. However, if both dipoles are excited equally (we assume this is the case under above band excitation) the visibility is modified as shown in eq. \ref{eq:vis_counts_full}, which could explain the fact that different emitters show different interference visibilities and could provide a way to roughly estimate the lateral offset of the emitter. 

The spontaneous emission rate in the waveguide $\gamma_0$ is instead derived from the local density of optical states (LDOS) $\rho_0(r_0,\omega)$, for the TE0 waveguide mode:
\begin{equation}
    \gamma_0 = \frac{\pi\omega|\mathbf{d}|^2}{3\hbar\varepsilon_0} \rho_0(r_0,\omega),
\end{equation}
where the LDOS is obtained from the dyadic Green's function $\overleftrightarrow{\mathbf{G}}(r,r',\omega)$ as follows:
\begin{equation}
    \rho_0 = \frac{6\omega}{\pi c^2} \left[\mathbf{p} ^\intercal\text{Im}\overleftrightarrow{\mathbf{G}}(r_0,r_0,\omega) \mathbf{p}\right]
\end{equation}
In a waveguide without mirrors, the Green's function is \cite{chen2010finite,hummer2013weak}:
\begin{equation}
    \overleftrightarrow{\mathbf{G}}(r,r',\omega) = \frac{c^2}{\omega v_g N} \mathbf{\bar{E}}(y,z)  \left[\mathbf{\bar{E}}^\intercal(y',z')\right]^\ast k G_0(x,x',\omega)
\end{equation}
with normalization factor $N = \int \varepsilon_r(y,z) |\mathbf{\bar{E}}(y,z)|^2 dydz$ and
\begin{equation}
    G_0(x,x',\omega) = \frac{i}{2k}e^{ik|x-x'|}
\end{equation}
being the 1D scalar Green's function.

The mirror produces an image dipole at a position $r''$ located at a distance $2L$ from the dipole position along the $x$-axis, which results in a total Green's function:
\begin{equation}
    \overleftrightarrow{\mathbf{G}}_\text{TOT}(r,r',\omega) = \overleftrightarrow{\mathbf{G}}(r,r',\omega) \pm |r_\text{T}| e^{i2\phi} \overleftrightarrow{\mathbf{G}}(r,r'',\omega),
\end{equation}
where the + holds for $x-$polarized dipoles and - holds for $y-$polarized dipoles due to the opposite reflection properties.  
The lifetime for a given dipole orientation is thus given by
\begin{equation}
\frac{\gamma_{x/y\phi}}{\gamma_{x/y0}} = \frac{\rho_{x/y\phi}}{\rho_{x/y0}} =  1 \pm |r_\text{T}| \frac{\text{Im}\left\{e^{i2\phi}G_0(0,2L,\omega)\right\}}{\text{Im}G_0(0,0,\omega)} 
\end{equation}
which results in the expression provided in \ref{eq:ldoschange}. We note that the change in lifetime is thus a direct manifestation of the change in LDOS. Importantly, this effect is very broadband, as $\omega$ does not play a significant role in the change of LDOS, in striking contrast with cavity-based approaches, where spectral tuning is essential to observe a change in the spontaneous emission rate.


\bibliography{bibliography}

\end{document}